\documentclass{aa}

\usepackage[varg]{txfonts}
\usepackage{graphicx}
\usepackage{multirow}
\graphicspath{{figures/}}
\usepackage{natbib}
\bibpunct{(}{)}{;}{a}{}{,}
\usepackage{hyperref}
\hypersetup{
        colorlinks=true,
        linkcolor=blue,
        citecolor=blue
}

\newcommand{\di}{{i\mkern1mu}}
\newcommand{\xigp}{\xi_\mathrm{g+}}
\newcommand{\rp}{r_\mathrm{p}}
\newcommand{\ud}{\mathrm{d}}
\usepackage[normalem]{ulem}

\begin{document}

\title{Intrinsic galaxy alignments in the KiDS-1000 bright sample: Dependence on colour, luminosity, morphology, and galaxy scale}
\titlerunning{Intrinsic galaxy alignments in the KiDS-1000 bright sample}

\author{Christos Georgiou \inst{1,2}\thanks{cgeorgiou@ifae.es}
    \and Nora Elisa Chisari \inst{1,3}
    \and Maciej Bilicki \inst{4}
    \and Francesco La Barbera \inst{5}
    \and Nicola R. Napolitano \inst{5,6,7}
    \and Nivya Roy \inst{8}
    \and Crescenzo Tortora \inst{5}
}
\institute{Institute for Theoretical Physics, Utrecht University, Princetonplein 5, 3584 CC, Utrecht, The Netherlands.
\and Institut de Física d’Altes Energies (IFAE), The Barcelona Institute of Science and Technology, Campus UAB, 08193 Bellaterra (Barcelona), Spain.
\and Leiden Observatory, Leiden University, Niels Bohrweg 2, NL-2333 CA Leiden, The Netherlands.
\and Center for Theoretical Physics, Polish Academy of Sciences, al. Lotników 32/46, 02-668 Warsaw, Poland.
\and INAF -- Osservatorio Astronomico di Capodimonte, Salita Moiariello 16, 80131 Napoli, Italy.
\and Department of Physics “E. Pancini”, University of Naples Federico II, Via Cintia, 21, 80126 Naples, Italy.
\and INFN, Sez. di Napoli, via Cintia, 80126, Napoli, Italy.
\and Department of Physics, Carmel College (Autonomous), Mala, Thrissur 680732, Kerala, India.
}
        
\date{Received <date> / Accepted <date>}

\abstract 
{
The intrinsic alignment of galaxies is a major astrophysical contaminant to weak gravitational lensing measurements, and the study of its dependence on galaxy properties helps provide meaningful physical priors that aid cosmological analyses. This work studied for the first time the dependence of intrinsic alignments on galaxy structural parameters. We measured the intrinsic alignments of bright galaxies, selected on apparent $r$-band magnitude $r<20$, in the Kilo-Degree Survey (KiDS). Machine-learning-based photometric redshift estimates are available for this galaxy sample that helped us obtain a clean measurement of its intrinsic alignment signal. We supplemented this sample with a catalogue of structural parameters from S\'ersic profile fits to the surface-brightness
profiles of the galaxies. We split the sample on galaxy intrinsic colour, luminosity, and S\'ersic index, and we fitted the non-linear linear alignment model to galaxy position--shape projected correlation function measurements on large scales. We observe a power-law luminosity dependence of the large-scale intrinsic alignment amplitude, $A_\mathrm{IA}$, for both the red and high-S\'ersic-index ($n_s>2.5$) samples, and find no significant difference between the two. We measure an $\sim1.5\sigma$ lower $A_\mathrm{IA}$ for red galaxies that also have a S\'ersic index of $n_s<4$ compared to the expected amplitude predicted using the sample's luminosity. We also probe the intrinsic alignment of red galaxies as a function of galaxy scale by varying the radial weight employed in the shape measurement. On large scales (above 6 Mpc/$h$), we do not detect a significant difference in the alignment. On smaller scales, we observe that alignments increase with galaxy scale, with outer galaxy regions showing stronger alignments than inner regions. Finally, for intrinsically blue galaxies, we find $A_\mathrm{IA}=-0.67\pm1.00$, which is consistent with previous works, and we find alignments to be consistent with zero for the low-S\'ersic-index ($n_s<2.5$) sample. 
}


\maketitle
\nolinenumbers

\section{Introduction}
\label{sec:Introduction}

The orientation of galaxies in the Universe, usually probed statistically with wide astronomical imaging surveys, has gathered significant attention in the last few decades (see \citealt{IAreviewJoachimi} for a historical review). It is understood that galaxy shapes are not randomly oriented in the Universe, instead, they have an imprint of the tidal gravitational field that can be studied through correlations involving galaxy shapes \citep[e.g.][]{reviewTroxel,IAGuide}. This intrinsic alignment (IA) of galaxies establishes galaxy shapes as a tracer that can reveal cosmological information, such as the imprint of baryon acoustic oscillations \citep[see e.g.][]{IABAODGC}, primordial gravitational waves \citep{Schmidt14,Chisari14}, primordial non-Gaussianity \citep{Schmidt15}, parity-breaking \citep{Biagetti20}, and the growth of the structure \citep{Taruya20,Okumura22,Okumura23}, among other applications. 

Besides the cosmological information available in galaxy intrinsic-shape statistics, another important reason to study the phenomenon is its implications for measurements of weak gravitational lensing. The apparent distortion of light bundles from the intervening matter distribution produces correlations of observed galaxy shapes that encode cosmological information \citep{WLreview}. Consequently, cosmic shear -- the weak gravitational lensing caused by the large-scale structure of the Universe -- is one of the main science drivers of prominent astronomical survey missions. Examples include ongoing or recently completed Stage-III surveys such as the Kilo-Degree Survey \citep[KiDS,][]{KiDS}, the Dark Energy Survey \citep[DES,][]{DES}, the Hyper Suprime-Cam Subaru Strategic Program \citep[HSC,][]{HSC}, as well as starting or upcoming Stage-IV missions such as Euclid \citep{Euclid}, the Vera C. Rubin Observatory LSST \citep{LSST}, or the Roman Space Telescope \citep{Roman}. Since galaxy shapes are not intrinsically random, the IA signal needs to be carefully considered when measuring weak gravitational lensing in order to arrive at accurate, unbiased cosmological results \citep[see e.g.][]{IAreviewKirk}. 

The need for mitigation of the IA signal's contamination of weak gravitational lensing measurements has driven much of the development in the field. Advancements in the theoretical modelling now go beyond the  linear alignment (LA) model of \cite{LA} and \cite{Hirata} and the phenomenological non-linear linear alignment model \citep[NLA,][]{bridleking} -- which substitutes the linear matter power spectrum for the non-linear counterpart in the LA model. State-of-the-art models consider the full expansion of biased tracers in the matter density field with standard perturbation theory \citep{Blazek19,LagrangianIA} or effective field theory \citep{Vlah1, Vlah2}. These more complex models have been shown to adequately describe the quasi-linear regime of IA correlations in simulated data \citep{Bakx} and are necessary to reach the stringent accuracy requirements of future cosmic shear surveys on the scales of interest \citep{Paopiamsap}. Hybrid Lagrangian extensions also serve to reduce the number of free parameters of these models and extend their validity regime \citep{Maion24}. On even smaller scales, comparable to the sizes of dark-matter haloes, the only theoretical description available comes from halo models of intrinsic alignments (\citealt{SchneiderBridle, Fortuna}; see \citealt{halomodelreview} for a review).

At the same time, the IA signal has been extensively studied in observational data, enabled by the wide-field imaging surveys of this century. Galaxy samples have been typically split in two populations: intrinsically red galaxies (which are generally elliptical and pressure supported) and blue galaxies (which are generally spiral and rotationally supported), since the IA signal is believed to depend strongly on the galaxy support mechanism. Intrinsic alignments have been detected for samples of intrinsically red galaxies \citep{Hirata07,Joachimi, Singh15, GAMAJohnston, KiDSLRG, Samuroff, UNIONS} with high significance, confirming the validity of the linear alignment model on large scales. On quasi-linear scales, the accuracy of IA measurements have not yet revealed any deviations from the effective description of the NLA model. For blue or emission-line galaxies the picture is more complicated; no detection has been made with the statistics that are  useful for quantifying IA contamination to weak lensing \citep{Mandelbaum11, Tonegawa18, GAMAJohnston, PAUSJohnston, Samuroff, HSCblueIA}, but such galaxies have been shown to align in some numerical simulations, albeit to different degrees of significance \citep{Chisari15,Tenneti16,Kraljic,Samuroff21,Delgado23}. Intriguingly, some simulations propose that their rotational axis aligns parallel to the separation vector, which would mean their shape alignment signal has the same sign as the cosmic shear signal (see e.g. \citealt{Chisari15} and \citealt{Kraljic}).
Consequently, the impact of blue galaxy alignments on cosmic shear measurements remains an open question.

The importance of IA in cosmology from weak gravitational lensing necessitates the development and testing of our models for IA through observations. This is especially important since the samples of galaxies that are used in cosmic shear studies differ significantly from those where intrinsic alignments can be studied directly and robustly. Therefore, understanding the general dependence of the IA signal on galaxy properties is crucial. 

It is understood that the large-scale alignment amplitude of red galaxies scales with their luminosity \citep{Joachimi,Singh15,KiDSLRG}. Recently, the universality of this scaling has been called into question, with red galaxies occupying different regions of the intrinsic colour -- absolute magnitude diagram shown to follow inconsistent scaling relations \citep{Samuroff}. On theoretical grounds, the intrinsic colour of a galaxy does not directly determine its alignment mechanism; rather, it is the mechanism of internal dynamical support with galaxies split into pressure-supported and rotationally supported types \citep{LA,IAreviewKiessling,Ghosh}. This distinction in galaxy populations might manifest more clearly with measurements of the galaxy's morphology rather than its intrinsic colour alone. 

In this work, we aim to test the universality of the large-scale IA amplitude scaling with galaxy luminosity using data from KiDS. We used a bright sub-sample of the full survey, ensuring the high quality of the estimated photometric redshifts (photo-$z$s), together with a catalogue of S\'ersic surface-brightness profile fits, providing morphological information of the galaxies. We also studied the alignment of blue galaxies and obtain upper bounds on their signal. Lastly, we utilised the flexibility of our shape measurement method to study the degree to which the IA signal depends on the scale of the galaxy being probed, a dependence that has been shown in \cite{Singh16} and \cite{Georgiou_groups}. Such a dependence can influence the choice of intrinsic alignment priors, and it might also have further applications with regard to IA mitigation \citep{Leonard18} and cosmology \citep{ChisariMulti}.

In Sect. \ref{sec:Data}, we introduce the data sets used in this work, the shape measurement procedure, and the morphological classification strategy. Section \ref{sec:Methodology} describes our methodology for measuring and modelling the relevant correlations and their covariance matrix. Section \ref{sec:Results} presents our results, and conclusions are given in Sect. \ref{sec:Conclusion}.
Throughout this work, we adopted a flat $\Lambda$-Cold-Dark-Matter ($\Lambda$CDM) cosmology with the best-fit cosmological parameters from the {\it Planck} 2018 analysis \citep{Planck18}, from the \texttt{Plik} likelihood, and we use $\Omega_m=0.315,\Omega_b=0.0493, h=0.673, \sigma_8=0.811$ and $n_s=0.965$\footnote{In the rest of the manuscript, $n_s$ is used to describe the S\'ersic index and not the spectral index of the primordial power spectrum.}.

\section{Data}
\label{sec:Data}

Direct measurements of the intrinsic alignment of galaxies require a galaxy catalogue containing shape and redshift information from which physically associated galaxies can be identified, typically in pairs. The galaxies for which shapes have been successfully measured constitute the shape sample. The highest IA signal is measured when this shape sample is correlated with positions of galaxies, which is called the density sample. These two samples do not need to contain the same galaxies, but they have to overlap in real space. When measuring real-space correlation functions, as we did in this work, a sample of random points that follow the angular and radial (in redshift) selection function of the two galaxy samples is also required. In this section, we detail the characteristics of the shape, density, and random samples employed in this work. 

\subsection{Kilo-Degree Survey}
\label{sec:Data_KiDS}

We studied the intrinsic alignment of galaxies using data from the Kilo-Degree Survey \cite[KIDS,][]{KiDS}. KiDS is a wide-field imaging survey carried out using the European Southern Observatory (ESO) VLT Survey Telescope. The survey covers four optical broad bands, $ugri$, and survey operations are now completed, with a total coverage of $1,347$ deg$^2$ \citep{KiDSDR5}. The recent data releases also combine the optical data with photometry from the VISTA Kilo-degree INfrared Galaxy (VIKING) survey \citep{Viking}, resulting in nine-band $ugriZYJHK_s$ photometry. KiDS was designed with weak gravitational lensing as its main scientific goal, with $r$-band observations restricted to the best dark-time night-sky conditions. As a result, the $r$-band images exhibit very high quality, with a limiting magnitude depth of 24.8 (at 5$\sigma$) and a median seeing of $0.7''$. This makes it an excellent choice for measuring galaxy shapes and extracting their intrinsic alignment signal.

In this work, we used the KiDS `bright' galaxy sample\footnote{\url{https://kids.strw.leidenuniv.nl/DR4/brightsample.php}} \citep{KiDS_Bright}. This sample is a high-quality subset of the full KiDS data, which contains galaxies with an apparent $r$-band magnitude of $r\leq20$ and was produced using the fourth KiDS data release \citep[DR4,][]{KiDS_DR4}. The sample selection was chosen to mimic, as closely as possible, galaxies from the equatorial fields of the Galaxy And Mass Assembly (GAMA) survey \citep{GAMA}, a highly complete spectroscopic survey. The third data release of the GAMA survey (labelled GAMA II) achieves $98.5\%$ completeness in these fields -- which cover 180 deg$^2$ and fully overlap with the KiDS footprint -- down to a limiting magnitude of Petrosian $r$-band $r_\mathrm{petro}\leq19.8$ \citep{GAMAII} based on the Sloan Digital Sky Survey (SDSS) DR7 photometry \citep{SDSSDR7}. This makes it a well-suited calibration sample for acquiring photo-$z$ estimates for the KiDS bright galaxies. These photo-$z$s are obtained using the neural-network-based redshift estimation software \textsc{ANNz2} \citep{ANNz2} with GAMA galaxies serving as the training sample, resulting in a high-quality photo-$z$ sample with a mean bias of $\sim5\times10^{-4}$ and a photo-$z$ scatter of $\sim0.018(1+z)$ quantified as scaled median absolute deviation (SMAD). We restricted the sample to galaxies outside masked regions and also with photo-$z$ in the range of $0.05\leq z_\mathrm{phot}\leq0.5$, as recommended in \cite{KiDS_Bright}, where the photo-$z$ performance is optimal. 

\begin{figure*}
    \centering
    \includegraphics[width=0.95\textwidth]{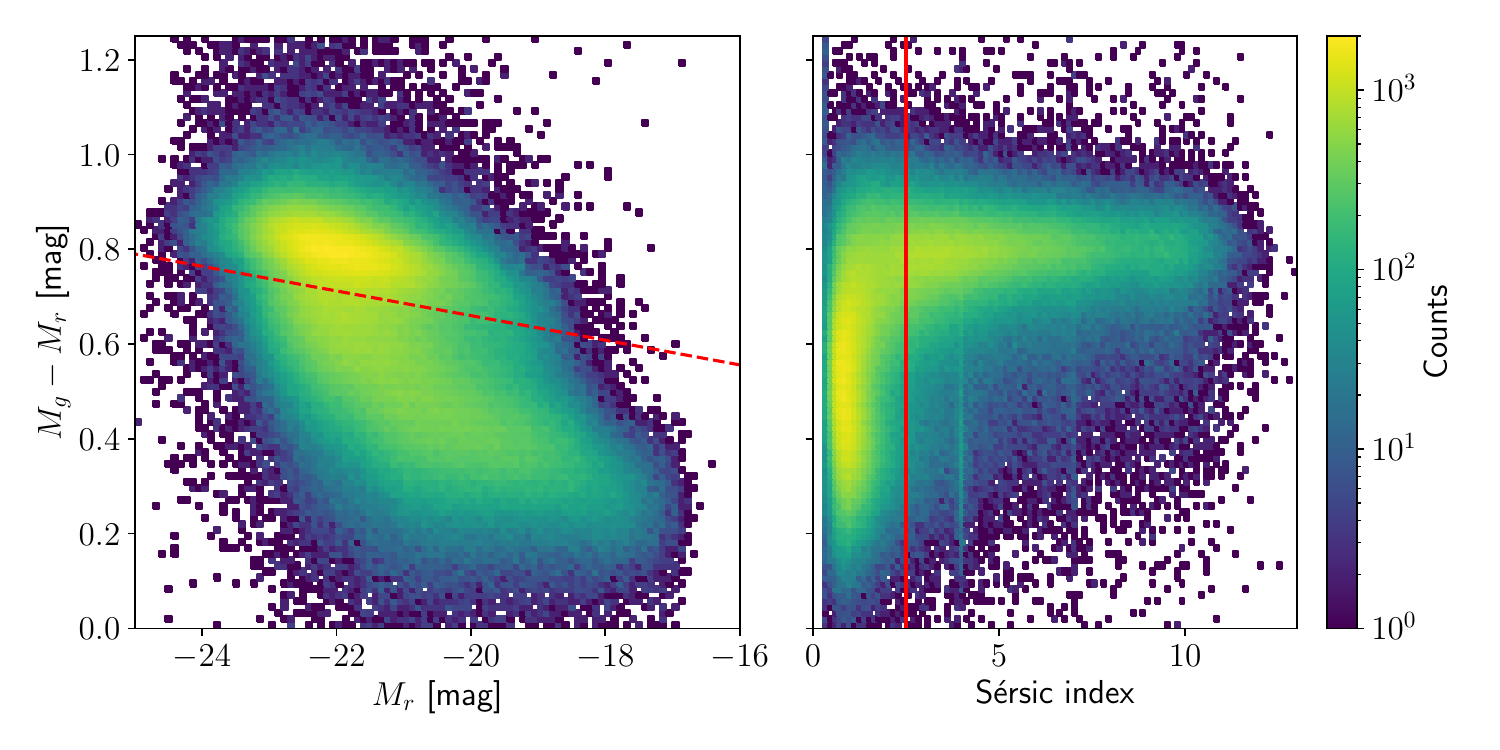}
    \caption{Two-dimensional histograms of properties of galaxies in final sample of KiDS bright catalogue. \emph{Left:} Distribution of rest-frame $g-r$ colour ($y$-axis) against absolute magnitude in the $r$ band ($x$-axis). The dashed red line shows our criteria for splitting the sample in 
    intrinsically red and blue galaxies. \emph{Right:} Same as before, but with the S\'ersic index, $n_s$, in the $x$-axis (see Sect. \ref{sec:Data_morphology} for details). The solid red line indicates our $n_s=2.5$ selection for high- and low-S\'ersic-index galaxies. }
    \label{fig:colour_mag_ns}
\end{figure*}

Accompanying the released KiDS bright sample is a catalogue of galaxy stellar masses and rest-frame absolute magnitudes produced using the spectral-energy-distribution (SED) template-fitting software \textsc{LePhare} \citep{LePhare}. We made use of this catalogue to acquire rest-frame colour and luminosity information for our galaxy sample, allowing us to divide galaxies into intrinsically red and blue populations. The distribution of galaxy rest-frame colour (computed using the absolute magnitudes, $M_g$ and $M_r$, in filters $g$ and $r$) compared to $r$-band absolute magnitude and the red and blue galaxy selection criteria can be seen in the left panel of Fig. \ref{fig:colour_mag_ns}; it is given by
\begin{equation}
    M_g-M_r=0.14-0.026M_r\,.
    \label{eq:red_blue_selection}
\end{equation}
The absolute magnitudes in the $g$ and $r$ bands in the above equation have also been corrected for the finite apertures used as input for \textsc{LePhare} and the different value of the dimensionless Hubble parameter, $h$, assumed in this work \citep[see Appendix C in][]{KiDS_Bright}. We also removed galaxies with inferred stellar masses outside the range of $7<\log_{10}(M/M_\odot)<12$ to avoid galaxies that might have problematic template fits, though this only affects $2,511$ galaxies. In total, all our cuts remove $20.8\%$ of the sample, which consists of $980,837$ galaxies. 

\subsection{DEIMOS galaxy shapes}
\label{sec:Data_shapes}

Galaxy shapes for the majority of the bright galaxy sample are not obtained with the standard KiDS pipelines due to limitations in the implementation of the shape-measurement pipeline, which prioritises faint galaxies. Here, we describe the galaxy shape catalogue that we produce for the KiDS bright sample using DEIMOS \citep{DEIMOS}, a galaxy surface-brightness moment-based shape-measurement method. The methodology here is identical to the one presented in \cite{Georgiou_DEIMOS}, and we refer the reader to that work for additional details of the DEIMOS implementation.

The ellipticity of galaxies is obtained via a combination of the second-order galaxy moments, which is given by
\begin{equation}
    \epsilon_1+\di\epsilon_2=\frac{Q_{20}-Q_{02}+\di2Q_{11}}{Q_{20}+Q_{02}+2\sqrt{Q_{20}Q_{02}-Q_{11}^2}}\,,
    \label{eq:ellipticity}
\end{equation}
where $\epsilon\equiv \epsilon_1+\di\epsilon_2$ is the galaxy ellipticity and $Q_{ij}$ are the (unweighted) moments of the galaxy surface brightness, $G(\mathbf{x})$, with the coordinate vector $\mathbf{x}=(x_1,x_2)$ given as 
\begin{equation}
    Q_{ij}=\int \ud\mathbf{x}\, G(\mathbf{x})x_1^i x_2^j \,.
    \label{eq:moments}
\end{equation}
Astronomical observations are sensitive to distortions due to optics and, for Earth-based observations, the atmosphere. These distortions are characterised by the point-spread function (PSF), which needs to be deconvolved from the observed moments to obtain the true galaxy moments. At the centre of the DEIMOS method is this deconvolution, which can be done analytically if the PSF moments are also known, up to the same order as the galaxy moments. 

Additionally, the presence of noise in astronomical images prevents the measurement of unweighted galaxy moments, making it necessary to employ a weighting scheme. A common choice, which we adopted here, is to use a multivariate Gaussian weighting function \citep{BernsteinJarvis} whose ellipticity and centroid is matched per galaxy. The size of this weight function, $r_\mathrm{wf}$, was fixed per galaxy and is related to the galaxy's isophotal radius, $r_\mathrm{iso}$, measured at $3\sigma$ above the noise background. Our fiducial sample was generated with $r_\mathrm{wf}=r_\mathrm{iso,}$ but we also explore other choices in Sect. \ref{sec:Results_weight}. The de-weighted moments are then recovered via a Taylor expansion, implemented to the fourth order in our case. 

The PSF was modelled using a shapelets expansion described in Appendix A1 of \cite{PSF_model}, and the second-order moments of the PSF were computed from this model at the position of each galaxy for the deconvolution. For the galaxy sample where shapes are necessary, we only used those that pass several quality control flags, ensuring the correct behaviour and convergence of the measured ellipticity. Lastly, we corrected the output ellipticity measurements for multiplicative bias, $m$, according to $\epsilon_\mathrm{true}=(1+m)\epsilon_\mathrm{observed}$, with values derived from \citet{Georgiou_groups} and summarised in Table \ref{tab:m-bias}. Note that additive bias is consistent with zero.

\begin{table}
    \caption{Multiplicative bias corrections for our shape measurements at different weight function sizes. }
    \label{tab:m-bias}
    \centering
    \begin{tabular}{c c}
        \hline\hline
        $r_\mathrm{wf}/r_\mathrm{iso}$ & $m$ \\
        \hline
        0.5 & 0.021 \\
        1.0 & -0.004 \\
        1.5 & -0.011 \\
        \hline
    \end{tabular}
\end{table}

\subsection{Galaxy morphology}
\label{sec:Data_morphology}

In order to study the dependence of galaxy intrinsic alignments on their morphology, we used an updated catalogue of galaxy structural parameters, which was first described in \cite{Roy} but are now applied to KiDS-DR4. This catalogue was generated by fitting a PSF-convolved 2D S\'ersic galaxy surface-brightness profile in the $r$-band KiDS DR4 images. The galaxy sample for this morphology catalogue was selected from the KiDS photometry catalogue by setting the star/galaxy separation flag $\texttt{CLASS\_STAR}=1$. In addition, the sample is restricted to galaxies with a high signal-to-noise ratio by requiring $1/\texttt{MAGAUTO\_ERR}>40$ in order to ensure accurate determination of the structural parameters. This corresponds to a completeness limit of $r<20.5$, which is deeper than that of the KiDS bright catalogue.

The estimation of galaxy structural parameters was done using the \textsc{2DPHOT} algorithm \citep{2dphot}. This algorithm performs PSF-convolved 2D S\'ersic profile fits to the galaxy surface brightness. The PSF correction is crucial for obtaining accurate structural parameters, and in \textsc{2DPHOT} this was done by selecting two or three (depending on distance) high-confidence stars that are close to the modelled galaxy and modelling the PSF 2D profile by fitting two Moffat profiles to those star images. The resulting structural parameters are the effective semi-major axis, $R_m$, central surface brightness, $I_0$, and S\'ersic index, $n_s$, of the best-fit S\'ersic profile, together with the axis ratio, $q$, position angle, photometric centre, and local background of each galaxy. From these parameters, we also obtain the circularised effective radius, $R_e=\sqrt{q}R_m$, and the apparent total modelled magnitude, $m_T$ (see \citealt{Roy} for further details on this procedure). 

The resulting galaxy morphology catalogue was further cleaned by requiring a sensible reduced chi-square value of the 2D profile fit, with $\chi^2_\mathrm{red}<10$. We also rejected galaxies that exhibit a very large effective radius, requiring $R_e<10$ arcsec, as the structural parameters of these galaxies are usually affected by large uncertainties. Lastly, we expect the modelled apparent magnitudes to be fairly close to the measured ones, and we required $|m_T-m_r|<1$. With this selection, we were left with $\sim4$ million galaxies. After spatial matching of the structural parameter catalogue with the KiDS bright sample, the resulting catalogue had 779,780 galaxies. We used this reduced sample only when selecting from the S\'ersic index and kept the larger, original, bright sample otherwise.

The parameter we are most interested in for this analysis is the S\'ersic index, $n_s$, the value of which describes the steepness of the central surface-brightness profile. A low value of $n_s$ describes a smooth galaxy-brightness profile that extends further out, while a high value describes a profile more concentrated in the centre. Because of this, spiral galaxies are roughly described by a S\'ersic profile with $n_s=1$ (exponential profile), while elliptical galaxies are described with $n_s=4$ (de Vaucouleurs profile). Hence, the S\'ersic index might serve as a proxy to study the alignment of rotationally supported and pressure-supported galaxies, which is one of the goals of this work. 

In the right panel of Fig. \ref{fig:colour_mag_ns} we show the 2D histogram of the rest-frame colour and S\'ersic index of galaxies from the structural parameter catalogue matched with the KiDS bright catalogue. We notice that galaxies can be divided into two distinct populations. On the top of the figure's right panel, we find galaxies that are intrinsically redder ($g-r>0.7$) and have a wide range of $n_s$ values with most of the galaxies having a value close to $n_s=4$. On the left side of the panel, we see a clump of galaxies that are intrinsically bluer and have a much lower value of $n_s$, with most of them being close to $n_s=1$. This relation between galaxy colour and S\'ersic index is well established from previous surveys, for example \citet{Blanton03}. In \citet{Roy} it was shown that a split at $n_s=2.5$ best divides the galaxy population into spheroids and disks, and we adopted this value to split the sample into high- and low-S\'ersic-index galaxies.

\subsection{Galaxy And Mass Assembly}
\label{sec:Data_GAMA}

So far, we have detailed our sample of galaxies for which shapes have successfully been measured. In this section, we describe the galaxy sample acting as a density tracer. The first consideration would be to use the full KiDS bright sample as the density tracer, especially since we are not interested in measuring and interpreting the galaxy bias (i.e. the relation between the galaxy and the matter overdensity), but instead we treat it as a nuisance (see Sect. \ref{sec:Methodology}). However, the galaxy clustering measurement used to infer the linear galaxy bias is very sensitive to the accuracy of the catalogue of random points used to normalise the measurement \citep{randoms}. This sensitivity is not as severe in the projected galaxy position--shape correlation, which we measured to infer the intrinsic alignment signal, since it is less dependent on the position of galaxies and also has a much lower signal-to-noise ratio (S/N). Generating a random sample for a photometrically selected flux-limited sample presents additional challenges \citep{PAUSJohnston}.

Given the challenge of producing a random catalogue accurate enough for the clustering measurements, we chose to follow a different approach. We aimed to create a sample that matches GAMA II galaxies more closely than the full KiDS bright sample -- which contains some galaxies that are fainter than the GAMA Petrosian $r$-band limit \citep{KiDS_Bright}. To do this, we took advantage of the latest GAMA data release, GAMA III \citep{GAMAIII}, which assimilates data from the KiDS and VIKING surveys to provide more complete and homogeneous photometry of the GAMA galaxy sample \citep{GAMAIIIphot}. Consequently, a more precise magnitude limit can be constructed that matches both GAMA III and the KiDS bright galaxies, since the samples use the same photometry.

We selected GAMA III galaxies using the \texttt{gkvScienceCat} data management unit (DMU) version 02 \citep{GAMAIIIphot} and selected all objects with sample class $\mathrm{SC}\geq6$ corresponding to a sample with $90\%$ completeness down to an $r$-band magnitude of $19.77$. This sample excludes objects identified as ambiguous, stars, or masked. We were not interested in having a highly complete density sample and chose this SC cut with lower completeness but higher number density to increase the statistical power of our measurements. We supplemented this catalogue with local flow corrections (to account for flows in the nearby Universe) to the galaxy redshifts from the \texttt{DistancesFrames} DMU version 14 \citep{GAMAlocalflow}, and we used the \texttt{Z\_TONRY} column for our galaxy redshifts. 

\begin{figure}
    \centering
    \includegraphics[width=0.47\textwidth]{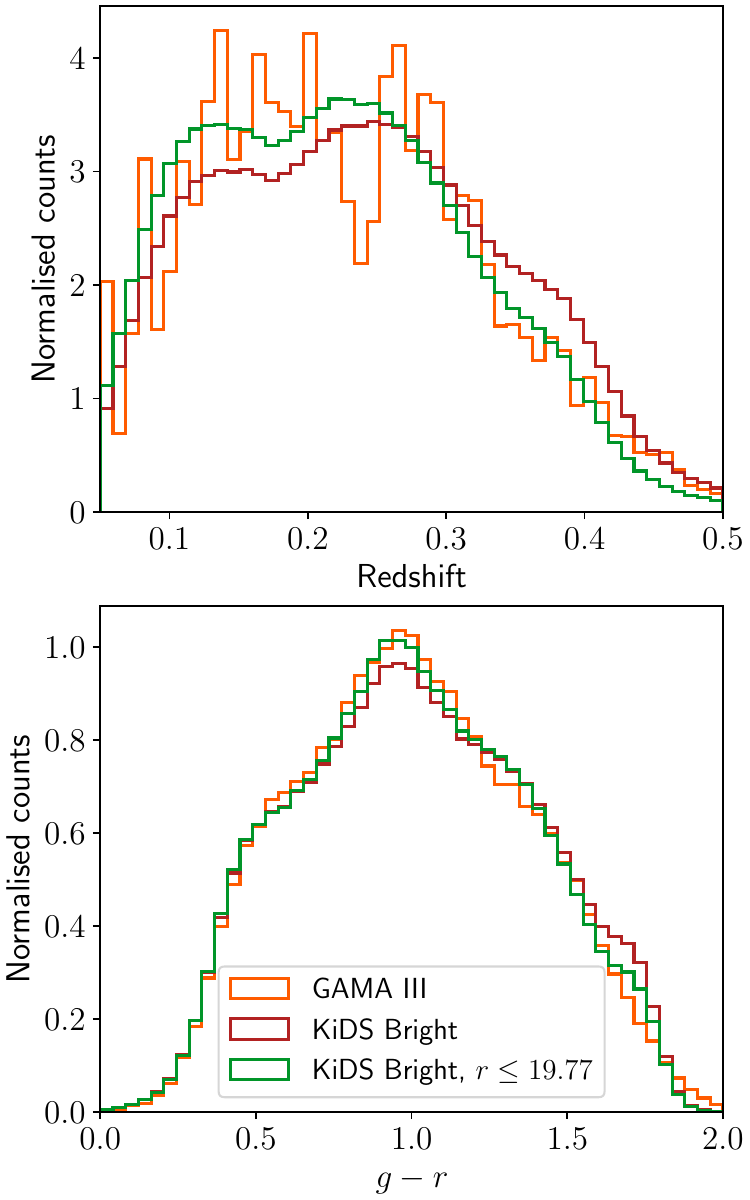}
    \caption{Redshift (top) and observed colour distribution (bottom) of the GAMA III (orange) and KiDS bright (red) samples. Limiting the KiDS bright sample to $r\leq19.77$ yields a clustering sample (green) that better matches the properties of the GAMA III sample.}
    \label{fig:redshift_colour}
\end{figure}

We used the resulting galaxy catalogue, restricted to the equatorial regions, to measure the clustering and infer the linear galaxy bias of the sample. The advantage of doing this is that we were able to use the randoms DMU specifically built for GAMA galaxies from \cite{GAMArandoms}\footnote{Note that these randoms were built for GAMA II, but they should be representative of the selection function of the GAMA III sample as well since the target selection was not changed, particularly in the equatorial regions.}. We then restricted the KiDS bright sample to $r\leq19.77$ and used it as our density sample in our projected galaxy position--shape correlations, and we assume the linear galaxy bias is the same as that for the GAMA III clustering measurement. We confirmed the selection produces a KiDS bright density sample that is similar to the GAMA III sample by examining the redshift and observed colour distribution, which is shown in Fig. \ref{fig:redshift_colour}. We see that the KiDS bright sample limited to an $r$-band magnitude of $r\leq19.77$ matches the observed $g-r$ colour and redshift distribution of the GAMA III sample better than that considering the full KiDS bright sample. These findings are consistent with what is seen in \citet{Jalan}, and the brighter cut implemented here is also expected to improve the quality of the photometric redshifts.

\subsection{Galaxy sub-samples}
\label{sec:Data_samples}

To examine the dependence of intrinsic alignments on colour, morphology, and both traits combined, we split the sample into several sub-samples, as detailed in Table \ref{tab:galaxy_samples}. Following the cuts presented in Fig. \ref{fig:colour_mag_ns}, we defined the red and blue samples in terms of their intrinsic rest-frame colour. The red sample was further split into five equipopulated luminosity bins using $L/L_0$, where $L_0$ is the luminosity of a galaxy corresponding to an absolute magnitude of $M_r = -22$. The luminosity bins contain approximately $80,000$ galaxies each. Note that these bins do not contain exactly the same number of galaxies due to galaxies being rejected from shape-measurement flags. We checked that these flags do not alter the sample characteristics seen in Table \ref{tab:galaxy_samples} by more than 1\% (and in most cases it is less than that). The mean redshift of the various sub-samples ranges from $0.15-0.36$.

Alternatively, we selected galaxies based on the S\'ersic index, placing a cut at $n_s>2.5$ to select elliptical galaxies as discussed in Sect. \ref{sec:Data_morphology}. We subdivided these galaxies further into the same five luminosity bins as for the red sample. Finally, we considered a joint colour and morphology cut by separating red galaxies into those that have $n_s<4$ and those that have $n_s>4$, which gives roughly equal sample sizes. This was done to further isolate dependence on morphology within our red sample. Distributions of galaxy properties for our different samples are presented in Appendix \ref{ap:distributions}.

In this analysis, we do not consider any evolution of the intrinsic alignment signal with redshift. Several studies have looked at this evolution and there does not seem to be a significant trend for the redshift baseline considered in this work \citep[e.g.][]{GAMAJohnston, Samuroff}. 

\begin{table}
    \label{tab:galaxy_samples}
    \caption{Different KiDS bright galaxy sub-samples considered in this work.}
    \centering
    \begin{tabular}{c c c c c}
        \hline\hline
        Selection & $\log(L/L_0)$ & $N_\mathrm{gal}$ & $\langle z\rangle$ & $\log\left(\langle L\rangle/L_0\right)$ \\
        \hline
        red & $(-\infty,-0.55]$ & 79584 & 0.16 & -0.74 \\
        red & $(-0.55,-0.33]$ & 81220 & 0.21 & -0.43 \\
        red & $(-0.33,-0.16]$ & 82043 & 0.26 & -0.24 \\
        red & $(-0.16,0.00]$ & 82498 & 0.31 & -0.08 \\
        red & $(0.00,+\infty)$ & 85354 & 0.36 & 0.16 \\
        \hline
        $n_s>2.5$& $(-\infty,-0.55]$ & 50464 & 0.15 & -0.77 \\ 
        $n_s>2.5$& $(-0.55,-0.33]$ & 52470 & 0.21 & -0.43 \\
        $n_s>2.5$& $(-0.33,-0.16]$ & 56212 & 0.26 & -0.24 \\
        $n_s>2.5$& $(-0.16,0.00]$ & 59040 & 0.31 & 0.08 \\
        $n_s>2.5$& $(0.00,+\infty)$ & 65429 & 0.36 & 0.16 \\
        \hline
        red, $n_s>4$ & - & 145297 & 0.28 & -0.09 \\
        red, $n_s<4$ & - & 203526 & 0.25 & -0.24 \\
        \hline
        blue & - & 385142 & 0.22 & -0.45 \\
        $n_s<2.5$ & - & 400786 & 0.22 & -0.42 \\
        \hline
        density & - & 675724 & 0.22 & -0.42 \\
        \hline
    \end{tabular}
    \tablefoot{Samples are split by colour, morphology, or both. We indicate the selection in luminosity ($\log(L/L_0)$), the number of galaxies in each bin ($N_{\rm gal
    }$), the mean redshift of the sub-sample ($\langle z \rangle$), and the logarithmic mean luminosity of the galaxies in the bin ($\log(\langle L\rangle/L_0)$).}
\end{table}

\section{Methodology}
\label{sec:Methodology}

We quantified the intrinsic alignment of galaxies by measuring the projected two-point correlation function between a galaxy-shape sample and a galaxy-position sample, the latter acting as a tracer of the density field. The projected correlation function between two fields, $\mathrm{a}$ and $\mathrm{b,}$ is given by
\begin{equation}
    w_\mathrm{ab}(\rp)=\int\ud z\,\mathcal{W}_\mathrm{ab}(z)\int_{-\Pi_\mathrm{max}}^{\Pi_\mathrm{max}}\ud\Pi\, \xi_\mathrm{ab}(\rp,\Pi,z)\,,
    \label{eq:wab}
\end{equation}
where $\rp$ and $\Pi$ are the projected and radial separations between tracers of fields $\mathrm{a}$ and $\mathrm{b,}$ and $\xi_\mathrm{ab}$ is the correlation function at redshift $z$. The correlation function $\xi_\mathrm{ab}$ is projected by integrating along the line-of-sight direction up to a maximum separation of $\Pi_\mathrm{max}$. The kernel function, $\mathcal{W}_\mathrm{ab}(z),$ depends on the comoving radial distance, $\chi(z)$, and the redshift distribution of the tracers, $n_\mathrm{a,b}(z)$, and is given by \citep{Mandelbaum11}
\begin{equation}
    \mathcal{W}_\mathrm{ab}(z)=\frac{n_\mathrm{a}(z)n_\mathrm{b}(z)}{\chi^2(z)\mathrm{d}\chi/\mathrm{d}z}\,\left(\int\ud z\frac{n_\mathrm{a}(z)n_\mathrm{b}(z)}{\chi^2(z)\mathrm{d}\chi/\mathrm{d}z} \right)^{-1}\,.
    \label{eq:W}
\end{equation}

The main observable of intrinsic alignment for this work is the correlation of galaxy shapes with galaxy positions. For any galaxy pair with a separation vector forming an angle, $\theta,$ with an arbitrary coordinate-system axis, the ellipticity of a galaxy, $\epsilon=\epsilon_1+i\epsilon_2$, can be decomposed in the tangential and cross-component with respect to the separation vector as
\begin{equation}
\begin{aligned}
    &\epsilon_+=\epsilon_1\cos(2\theta)+\epsilon_2\sin(2\theta)\\
    &\epsilon_\times = \epsilon_1\sin(2\theta)-\epsilon_2\cos(2\theta)\,,
\end{aligned}
\end{equation}
respectively. Here, we followed the convention that when the semi-major axis of the ellipse is parallel to the separation vector (indicating radial alignment), $\epsilon_+$ is positive. We computed the correlation function of galaxy positions and the tangential ellipticity component between galaxy pairs, denoted as $\xigp$. Note that if assuming parity is not violated in the Universe, the $\xi_\mathrm{g\times}$ correlation is expected to be zero. This serves as a null test for any systematic effects that are not accounted for in the shape-measurement method, such as imprecise PSF modelling. 

In the presence of uncertain redshift information, as is the case for our sample of galaxies with photometrically obtained redshifts, we need to take into account the probability distribution function of a galaxy's true redshift given the estimated redshift when modelling the correlation function. To do this, we first write the correlation function of galaxy shapes and positions following \cite{Joachimi}:
\begin{equation}
    \xigp^\mathrm{phot}(\rp,\Pi,z_m) = \int_0^{\infty}\frac{\ud \ell\,\ell}{2\pi}J_2\left(\ell\theta(\rp,z_m)\right)C_\mathrm{gI}\left(\ell|z_m,\Pi\right)\,,
    \label{eq:xigp}
\end{equation}
where $J_n(x)$ is the Bessel function of the first kind. The mean redshift, $z_m,$ at which the correlation function is estimated and the line-of-sight separations $\Pi$ are related to the redshifts of a galaxy pair, $z_1,z_2$, through
\begin{equation}
    \begin{aligned}
        &z_m=\frac{1}{2}(z_1+z_2)\,,\\
        &\Pi = \frac{c}{H(z_m)}(z_2-z_1)\,,
    \end{aligned}
    \label{eq:Pimax}
\end{equation}
where $c$ and $H$ are the speed of light and Hubble parameter, respectively, while the angle is given by $\theta=\rp\,\chi^{-1}(z_m)$. Equation \eqref{eq:xigp} relates the correlation function to the angular power spectrum, $C_\mathrm{gI}$, which is given by 
\begin{equation}
    \begin{aligned}
        C_\mathrm{gI}(\ell|z_1,z_2)=&\int_0^{\chi_\mathrm{hor}}\ud\chi'\frac{p_\mathrm{g}\left(\chi'|\chi(z_1)\right)p_\epsilon\left(\chi'|\chi(z_2)\right)}{\chi'^2}\,\times\\
        &P_\mathrm{gI}\left(k=\frac{\ell+0.5}{\chi'},z_m \right)\,.
    \end{aligned}
    \label{eq:CgI}
\end{equation}
Here, $p_{g,\epsilon}$ is the probability density function of a galaxy's true redshift given the observed one ($z_1$ or $z_2$), but expressed in terms of the comoving radial distance. Subscripts indicate the galaxy-position sample ($\mathrm{g}$) or the galaxy-shape sample ($\epsilon$), and the integral is carried out to the comoving radial horizon distance. We also used the \cite{Limber53} approximation.

The power spectrum of galaxy position and galaxy intrinsic shape, $P_\mathrm{gI}$, contains the statistical information of galaxy intrinsic alignments and is the relevant quantity we needed to model. We restricted our analysis to large enough scales where we were able to treat the tracer fields as related linearly to the matter density field, $\delta$. Consequently, we assumed linear galaxy bias, $b_\mathrm{g}$, with the galaxy density given by $\delta_\mathrm{g}=b_\mathrm{g}\delta,$ and we employed the linear alignment model \citep{LA, Hirata}, resulting in
\begin{equation}
    P_\mathrm{gI}(k,z)=-b_\mathrm{g}A_\mathrm{IA}\frac{\rho_\mathrm{crit}\Omega_\mathrm{m}\bar{C}_1}{D(z)}\,P_\delta(k,z)\,.
    \label{eq:PgI}
\end{equation}
The intrinsic alignment amplitude, $A_\mathrm{IA}$, is the quantity that describes the intrinsic alignment of galaxies on large scales, for a given set of cosmological parameters and modulated by the linear galaxy bias. A positive value of $A_\mathrm{IA}$ indicates galaxies orient their semi-major axis in the direction of the position of other galaxies, with higher values indicating a stronger statistical alignment. The amplitude, $\bar{C}_1$, is a normalisation factor, and a common choice is the value given in \cite{bridleking}, with $\bar{C}_1=5\times10^{-14}(h^2M_\odot/\mathrm{Mpc}^{-3})^{-2}$. Lastly, $\rho_\mathrm{crit}$ is the critical matter density, $D(z)$ is the growth factor normalised to 1 at $z=0,$ and $P_\delta$ is the matter power spectrum. The linear expansion described in \cite{Hirata} requires the linear matter power spectrum be used in Eq. \eqref{eq:PgI}. We followed \cite{bridleking} and used the non-linear matter power spectrum.

To obtain the probability density function of a galaxy's true redshift given its observed one, we computed the distribution of redshift error, $\delta z=(z-z_\mathrm{phot})/(1+z)$, and fitted a generalised Lorentzian distribution to it:
\begin{equation}
    p_\mathrm{g,\epsilon}(z,z_\mathrm{phot}) \propto \left(1+\frac{\delta z}{2\alpha s^2} \right)^{-a}\,.
\end{equation}
This distribution is shown to describe the photometric redshift errors of the sample very well, particularly in the long tails of the distribution \citep{KiDS_Bright}. We obtained the galaxy true redshift by spatially matching our galaxy sample to GAMA III galaxies (finding matches within 2 image pixels chosen to strike a balance between failing to find true matches and finding false ones), and then we determined the values of $\alpha$ and $s$ for each of our galaxy sub-samples (see Sect. \ref{sec:Data_samples}).

\begin{figure}
    \centering
    \includegraphics[width=0.47\textwidth]{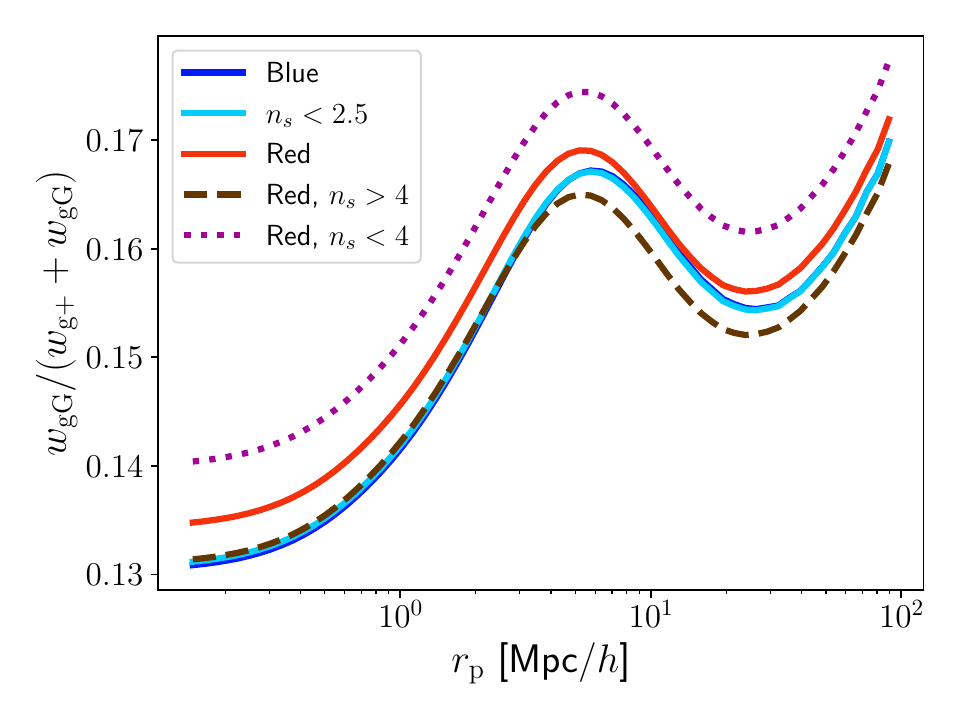}
    \caption{Contamination of weak gravitational lensing measurements in the intrinsic alignment measurements, expressed as a ratio of the lensing signal over the total signal.}
    \label{fig:lensing_contamination}
\end{figure}

When measuring correlations between galaxy shapes and positions, if galaxy pairs are separated by a large enough distance, the signal from the weak gravitational lensing of the background galaxy caused by the foreground galaxy can be non-negligible. This is especially common when galaxy distances are based on photometric redshifts, and galaxies need to be correlated to high $\Pi_\mathrm{max}$ in order to recover the intrinsic alignment signal that gets redistributed out to large values of $\Pi$ by the photo-$z$ uncertainty. The weak gravitational lensing contribution to the correlation function, $\xi_\mathrm{gG}$, has the same form as Eq. \eqref{eq:xigp}, but with the angular power spectrum substituted for
\begin{equation}
    \begin{aligned}
        C_\mathrm{gG}(\ell|z_1,z_2)=&\int_0^{\chi_\mathrm{hor}}\ud\chi'\frac{p_\mathrm{g}\left(\chi'|\chi(z_1)\right)q_\epsilon\left(\chi'|\chi(z_2)\right)}{\chi'^2}\,\times\\
        &P_\delta\left(k=\frac{\ell+0.5}{\chi'},z_m \right)\,,
    \end{aligned}
    \label{eq:CgG}
\end{equation}
where the lensing kernel for the galaxy shape sample is given by
\begin{equation}
    q_\epsilon(\chi,\chi_1) = \frac{3H_0^2\Omega_m}{2c^2}\,\frac{\chi}{a(\chi)}\int_\chi^{\chi_\mathrm{hor}}\ud\chi'p_\epsilon\left(\chi'|\chi_1\right)\,\frac{\chi'-\chi}{\chi'}\,,
\end{equation}
and $a(\chi)$ is the scale factor. In Fig. \ref{fig:lensing_contamination}, we show the ratio between the lensing contamination and the full signal, $w_\mathrm{gG}/(w_\mathrm{gG}+w_\mathrm{g+})$. The weak lensing contamination is stronger for samples with larger errors in their redshift distributions and for samples at higher redshift, with both effects playing an important role. In principle, lensing magnification will also impact our measured signal, but its effect is expected to be very small by comparison \citep{Samuroff}, especially over the relatively low redshift baseline of our galaxy sample. For this reason, we neglect its impact here.

The last ingredient in our modelling is the linear galaxy bias, $b_\mathrm{g}$, which we treated as a nuisance parameter. The most common way to constrain it is to measure the projected auto-correlation function of galaxy positions, the galaxy clustering. However, as mentioned in Sect. \ref{sec:Data_GAMA}, this measurement is highly dependent on the accuracy of the random points used in the Landy-Szalay estimator, which are even more challenging to construct for a flux-limited, photometric galaxy sample with photo-$z$ uncertainty \citep{PAUSJohnston}. Instead, we chose to select a sub-sample of the full KiDS bright sample to act as the density sample that better matches galaxies in the GAMA III sample. We then made the assumption that these two samples have the same linear galaxy bias and used the GAMA III sample to measure the galaxy clustering signal and determine $b_\mathrm{g}$. For a sample with precise redshift information, the projected galaxy clustering is equal to
\begin{equation}
    \begin{aligned}
        w_\mathrm{gg}^\mathrm{spec}(\rp)=&b_\mathrm{g}^2\int\ud z\,\mathcal{W}_\mathrm{ab}(z) \int\int\frac{\ud k_\perp \ud k_\parallel}{\pi^2}\frac{k_\perp}{k_\parallel}\sin(\Pi_\mathrm{max}k_\parallel)\,\times\\
        &J_0(k_\perp \rp)P_\delta\left(\sqrt{k_\parallel^2+k_\perp^2},z\right)\,,
    \end{aligned}
    \label{eq:wgg}
\end{equation}
with $k_\parallel$ and $k_\perp$ being the components of the Fourier wave-number vector that are parallel and perpendicular to the line of sight, respectively. 

A similar equation can be written for the position--shape correlation of a galaxy sample with accurate redshift information, as in the case of spectroscopic redshifts, where
\begin{equation}
    \begin{aligned}
        w_\mathrm{g+}^\mathrm{spec}(\rp)=&\int\ud z\,\mathcal{W}_\mathrm{ab}(z) \int\int\frac{\ud k_\perp \ud k_\parallel}{\pi^2}\frac{k_\perp^3}{(k_\parallel^2+k_\perp^2)k_\parallel}\sin(\Pi_\mathrm{max}k_\parallel)\,\times\\
        &J_2(k_\perp \rp)P_\mathrm{gI}\left(\sqrt{k_\parallel^2+k_\perp^2},z\right)\,.
    \end{aligned}
    \label{eq:wg+}
\end{equation}
We used this to validate our code that computes the photometric $w_\mathrm{g+}$ by ensuring the two predictions match when we use an extremely narrow distribution for $p_{\mathrm{g},\epsilon}(z,z_\mathrm{phot})$ in Eq. \eqref{eq:CgI}. 

\subsection{Estimators}
\label{sec:Methodology_estimators}

In this section, we describe how to connect the theoretical models of galaxy statistics, outlined in Sect. \ref{sec:Methodology}, to estimators obtained from observational data. We used the Landy-Szalay estimator \citep{LandySzalay} to measure the auto-correlation of galaxy positions, which is given by
\begin{equation}
    \hat{\xi}_\mathrm{gg}=\frac{DD-DR-RD+RR}{RR}\,.
    \label{eq:xiggest}
\end{equation}
The terms in the above equation denote measurements of galaxy-pair counts within a specific bin of galaxy-pair projected and line-of-sight separation, $\rp$ and $\Pi$. The variable $D$ is for a sample of galaxy positions taken from the density sample, and $R$ is for galaxy positions obtained from a sample of random data points that mimic the angular and radial selection function of sample $D$. We connected the measured correlation function to our observed, projected correlation function via
\begin{equation}
    \hat{w}_\mathrm{ab}(r_{\rm p})=\int_{-\Pi_\mathrm{max}}^{\Pi_\mathrm{max}}\,\ud\Pi\,\hat{\xi}_\mathrm{ab}(r_{\rm p},\Pi)\,,
    \label{wabest}
\end{equation}
where the integral is reduced to a Riemann sum over the $\Pi$ bins. For the galaxy clustering signal, measured from the GAMA III spectroscopic galaxies, we used $\Pi_\mathrm{max}=60$ Mpc$/h$ and binned $\Pi$ linearly in 30 bins. We used \texttt{TreeCorr}\footnote{\url{http://rmjarvis.github.io/TreeCorr} (version 4.3.3).} \citep{treecorr} to measure the correlation functions. We measured all correlations in 11 bins of $\rp$ logarithmically spaced from 0.1 to 60 Mpc$/h$.

For the measurement of the galaxy position--shape correlation, we followed \cite{Responsivity} and computed the estimator
\begin{equation}
    \hat{\xi}_\mathrm{g+}=\frac{S_+D-S_+R}{RR}\,.
    \label{eq:xigpest}
\end{equation}
Here, the terms on the nominator are the sum of the tangential ellipticity component of galaxy pairs that lie within a bin of $\rp$ and $\Pi$. This is given by
\begin{equation}
    S_+A=\sum_{i\neq j}\epsilon_+(i|j)/\mathcal{R}\,,
    \label{eq:SpD}
\end{equation}
where the galaxy $i$ belongs to the shape sample and the galaxy $j$ to the density or random sample, for $A=D$ and $R$, respectively. 
To connect the inferred intrinsic alignment amplitudes with their impact on measurements of weak gravitational lensing, we need to correct the ellipticity measurements for the responsivity\footnote{Note that the responsivity here differs from what was used in \citet{Responsivity}, because the ellipticity measure we used is the third flattening.} of the ellipticity estimate to the shear: $\mathcal{R}=\partial\epsilon/\partial\gamma\approx1-\epsilon_\mathrm{rms}^2$ \citep{Responsivity}. To speed up the computation of the correlations, we chose to use the pair counts $DR$ instead of $RR$ in the denominator of Eq. \eqref{eq:xigpest}, which has a negligible effect on the measurement \citep{GAMAJohnston}.

\subsection{Covariance estimation}
\label{sec:Methodology_covariance}

We used jackknife resampling to estimate the covariance of our measurements directly from the data. This method provides a way to estimate the covariance of a sample without the need to simulate the observations or develop an analytical description. It can be a noisy estimate of the covariance, but for large enough samples (large in area and number density in the case of astronomical observations) this noise does not significantly impact the estimation \citep{Hartlap}. However, it does not include contributions to the covariance from modes beyond the survey's window \citep[super-sample covariance,][]{SSC} for which an analytic description is not available in the case of intrinsic alignment measurements, and we chose to ignore it here given the relatively low S/N.

We split our sample by area into jackknife regions using a kmeans clustering algorithm in the sphere, \textsc{kmeans-radec}\footnote{\url{https://github.com/esheldon/kmeans-radec} (version 0.9.1).}. This ensures the different jackknife regions contain an approximately equal area and number density, avoiding the need to correct for that later on. By removing a jackknife patch labelled $a$ from the full sample and measuring the correlation function in this 'delete-one' jackknife sample we obtain a jackknife measurement $w^J$ of the projected correlation function (for $w_\mathrm{gg}$ or $w_\mathrm{g+}$). We then estimated the full sample's covariance matrix from $N$ jackknife sub-samples via
\begin{equation}
    C=\frac{N-1}{N}\sum_{a=1}^N(w^J-\bar{w})(w^J-\bar{w})^\mathrm{T}\,,
    \label{eq:covariance}
\end{equation}
where $\bar{w}$ is the average of all the delete-one jackknife measurements. The covariance estimated in this way is biased, and we applied a correction multiplying the covariance matrix by $(N-D-2)/(N-1)$ to correct this bias \citep{Hartlap}. Here, $N$ is the number of jackknife realisations and $D$ is the number of data points in our data vector. Lastly, we made sure to include the responsivity factor in the covariance from Eq. \eqref{eq:SpD}.

The number of jackknife patches plays an important role in the validity of the estimated covariance. We need jackknife patches that are large enough in area to sample the largest physical scales of interest, but we also need a large number of patches to avoid a very noisy covariance estimate. This is particularly challenging for low-redshift samples, where the area of a patch on the sky does not correspond to a particularly large physical separation between objects in that area (compared to high-redshift samples). 

We chose to use 40 jackknife regions that roughly correspond to patch sizes of 25 deg$^2$ or angular separations of 5.64 deg (if assumed circular). This is a good balance between the number of patches and scales we can analyse. The patch size corresponds to 60 Mpc$/h$ physical separation (our largest $\rp$ bin edge) at redshift of 0.28, which means galaxies at lower redshift inside the patch will be separated by less than 60 Mpc$/h$. A large fraction of galaxies fall below this redshift for all galaxy samples except the two most luminous ones, and for this reason we chose to ignore the highest $\rp$ bin in our analysis. For the second largest $\rp$ bin at 33.5 Mpc$/h$, this redshift is much lower: 0.14. The two lowest luminosity samples have approximately $40\%$ of their galaxies below this redshift, while for all other samples this fraction is much lower. Hence, we trust that the covariance for this bin is accurate.

\section{Results}
\label{sec:Results}

In this section, we present the results from measurements of the intrinsic alignment signal of galaxies in the sub-samples defined in Sect. \ref{sec:Data_samples}. We measured the position--shape two-point projected correlation function and the jackknife covariance. We also measured the position--position projected correlation of the GAMA III sample. We then jointly fitted the $w_\mathrm{gg}$ and $w_\mathrm{g+}$ measurements for each sub-sample to determine the linear galaxy bias and the intrinsic alignment amplitude, $A_\mathrm{IA}$. Since we used linear models for the galaxy bias and intrinsic alignments, we restricted our fits to scales of $r_p > 6$ Mpc$/h$. The best-fit linear galaxy bias --  which is the same in all our measurements since the sample for galaxy positions remains the same -- was determined to be $b_\mathrm{g}=1.23\pm0.09$. Model power spectra were computed using the \textsc{Core Cosmology Library}\footnote{\url{https://github.com/LSSTDESC/CCL}} \citep[CCL][]{CCL} version 3.1.2, and the fitting was done using the \textsc{scipy} python library \citep{scipy}. The non-linear matter power spectrum was obtained using \textsc{HaloFit} \citep{halofit}.

\begin{figure*}[ht!]
    \centering
    \includegraphics[width=0.95\textwidth]{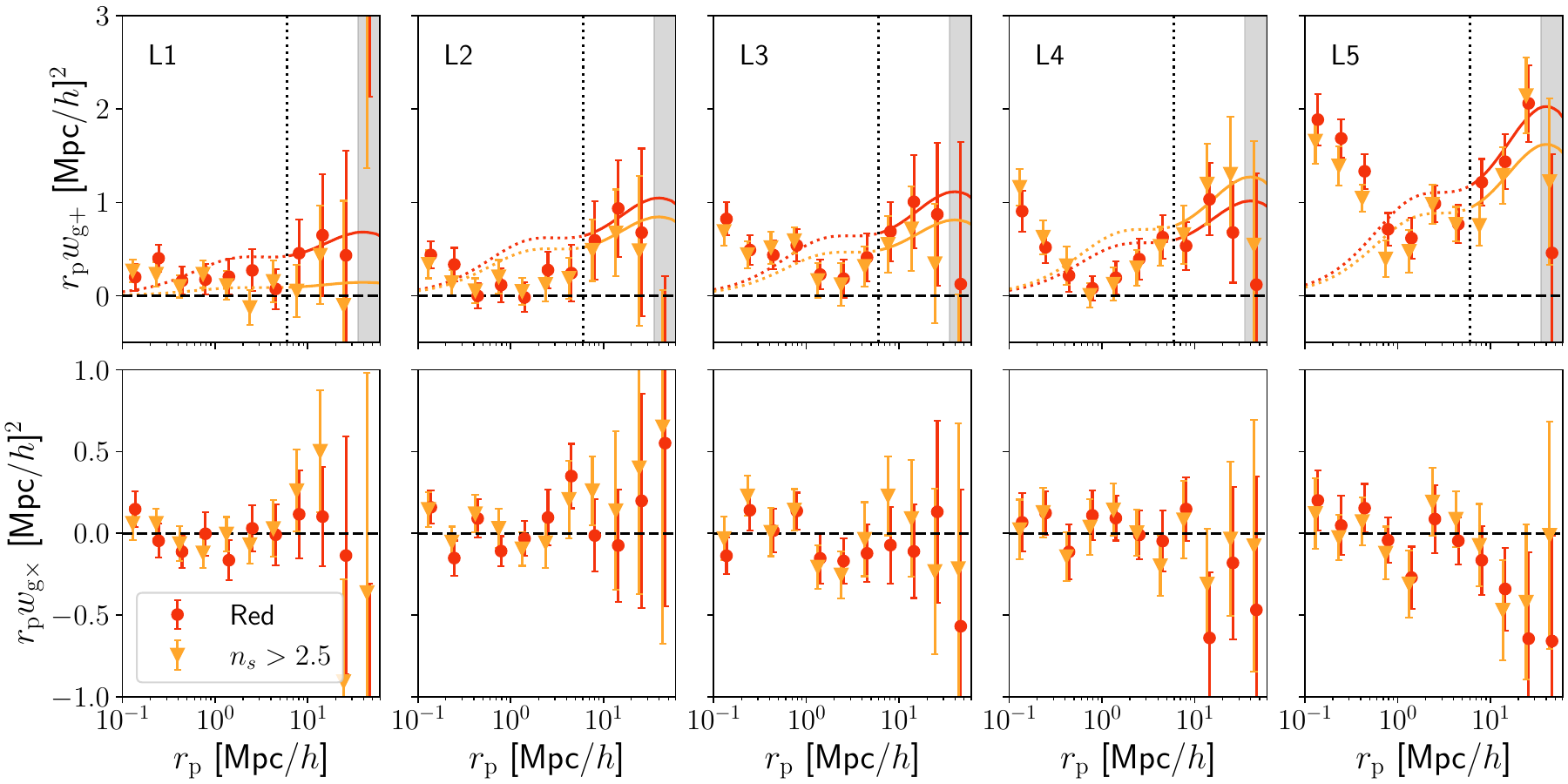}
    \caption{\emph{Top:} Projected position--shape correlation function measurements as a function of galaxy pair separation for the red (red circles) and high $n_s$ (orange triangles) galaxy sub-samples. The different luminosity sub-samples are indicated in the top left of each panel, with luminosity increasing from left to right. The best-fit NLA model is shown as solid lines of corresponding colour -- with dotted lines where the model is no longer valid, at $\rp<6$ Mpc$/h,$ which is also indicated with a vertical dotted black line. The grey area indicates the region where the covariance estimation is no longer accurate. Points are slightly offset for visual clarity.
    \emph{Bottom:} Projected correlation measurements between galaxy positions and the cross ellipticity component.}
    \label{fig:luminosity}
\end{figure*}

\subsection{Luminosity dependence of IA}
\label{sec:Results_luminosity}

We first focus on the well-studied dependence of the intrinsic alignment signal on the average luminosity of the galaxy sample. We measured $A_\mathrm{IA}$ for our intrinsically red sample split into five different luminosity bins (labelled L1 through L5 with increasing luminosity) and we repeated this for samples selected according to S\'ersic index, with $n_s>2.5$. We present the measurements together with the best-fit NLA model in Fig. \ref{fig:luminosity}. 

Firstly, we see that the position--cross ellipticity correlation (bottom panels) is consistent with 0 across all samples, which serves as a good null test for systematics in the shape-measurement method. Looking at the $w_\mathrm{g+}$ measurements, we see that galaxies of higher luminosity show a stronger alignment overall, a result consistent with several other studies in the literature. We do not see a significant difference in the measured correlations between the samples selected on colour (red) or morphology ($n_s$), given the measured uncertainty. 

\begin{table}
    \caption{Best-fit NLA amplitude and luminosity-scaling parameters.}
    \label{tab:A_IA}
    \centering
    \begin{tabular}{c c c c}
        \hline\hline
        Selection & $A_\mathrm{IA}$ & $A_0$ & $\beta$ \\
        \hline
        red (L1)& $2.44\pm1.41$ & \multirow{5}{*}{$5.95\pm0.49$} & \multirow{5}{*}{$0.68\pm0.18$}  \\
        red (L2)& $3.82\pm1.59$ & & \\
        red (L3)& $4.14\pm1.23$ & & \\
        red (L4)& $3.96\pm1.18$ & & \\
        red (L5)& $8.07\pm1.04$ & & \\
        \hline
        $n_s>2.5$ (L1)& $0.51\pm1.20$ & & \\
        $n_s>2.5$ (L2)& $3.04\pm1.30$ & & \\
        $n_s>2.5$ (L3)& $3.03\pm1.25$ & $5.11\pm0.43$ & $0.79\pm0.16$ \\
        $n_s>2.5$ (L4)& $4.98\pm1.61$ & & \\
        $n_s>2.5$ (L5)& $6.54\pm1.45$ & & \\
        \hline
        red, $n_s>4$ & $5.24\pm0.85$ & & \\
        red, $n_s<4$ & $1.12\pm1.18$ & & \\
        \hline
        blue & $-0.67\pm1.00$ & & \\
        $n_s<2.5$ &  $0.64\pm0.88$ & & \\
        \hline
    \end{tabular}
\end{table}

\begin{figure}
    \centering
    \includegraphics[width=0.47\textwidth]{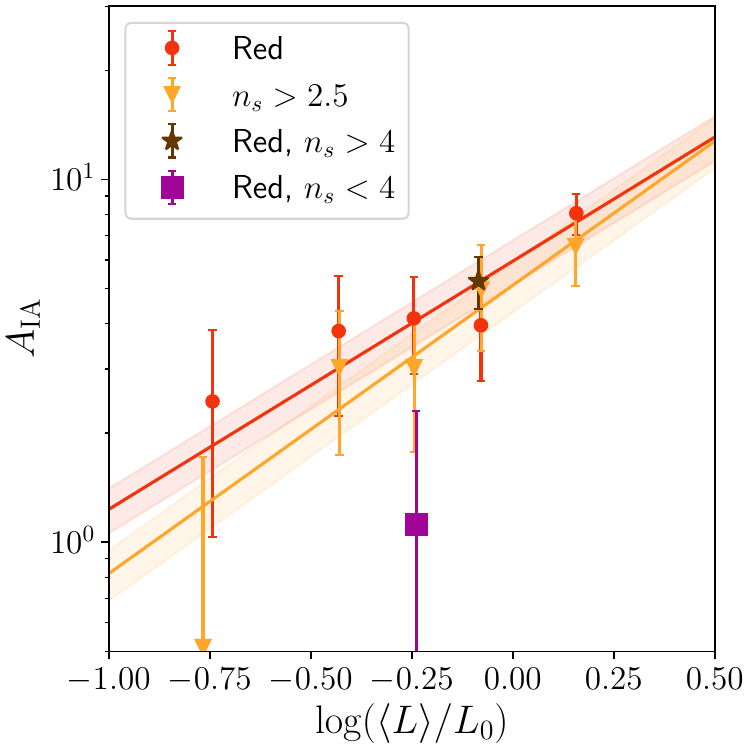}
    \caption{Best-fit intrinsic alignment amplitude of NLA fits to $w_\mathrm{g+}$ measurements from the different galaxy sub-samples, plotted against their average luminosity logarithm. The red or $n_s>2.5$ galaxy samples (red circles and orange triangles, respectively) are split into five luminosity bins, while the galaxy samples selected on both colour and $n_s$ (brown star or square) are not split in luminosity. A power-law luminosity-dependent $A_\mathrm{IA}$ is fitted to the intrinsically red and $n_s>2.5$ samples, plotted as red and orange solid lines, respectively, with the colour band indicating 1-$\sigma$ intervals of the best-fit parameters.}
    \label{fig:luminosityfit}
\end{figure}

The resulting best-fit NLA model is also seen in the figure, with the values of $A_\mathrm{IA}$ shown in Table \ref{tab:A_IA} and plotted more clearly in Fig. \ref{fig:luminosityfit} for the different samples. While for most of our measurements the $A_\mathrm{IA}$ of high-$n_s$-selected galaxies is lower than for red galaxies, this difference is small compared to the uncertainty in the derived $A_\mathrm{IA}$. We fitted the dependence of alignment amplitude on luminosity with a single power law,
\begin{equation}
    A_\mathrm{IA}(L)= A_0\left(\frac{\langle L\rangle}{L_0}\right)^\beta\,,
    \label{eq:luminosity}
\end{equation}
and we fitted for $A_0$ and a slope $\beta$. The best-fit values for these parameters are shown in Table \ref{tab:A_IA}, and the parameters are approximately within 2-$\sigma$ of each other's interval (however, they are not independent).

\subsection{Morphological dependence of red galaxies}
\label{sec:Results_morphology}

In the previous section we validated that the dependence of alignment amplitude on the galaxy sample's average luminosity is consistent if the samples are selected according to colour or morphology. This dependence has been observed so far for intrinsically red galaxy samples, but it is not yet clear if this dependence is universal. In \citet{Samuroff}, it was hinted that the colour--magnitude distribution of the galaxy sample might play a role in at least the amplitude of the luminosity dependence, $A_0$. Here, we aim to probe this question by selecting the intrinsically red galaxies in our sample and then splitting by morphology, to answer whether the structure of galaxies might affect this relation more strongly.

We measured $w_\mathrm{g+}$ and fitted for $A_\mathrm{IA}$ for red galaxies with $n_s>4$ and $n_s<4$, and the results are shown in Table \ref{tab:A_IA}. We see a large difference in the amplitude of the alignment signal, with the low S\'ersic index sample having a much lower amplitude than the high-$n_s$ sample. We then compare the measured value of $A_\mathrm{IA}$ to the expected one given the average luminosity of the samples. Those values are over-plotted in Fig. \ref{fig:luminosityfit}. We see that $A_\mathrm{IA}$ for the high-$n_s$ red sample is consistent with the amplitude we expect from luminosity dependence alone. However, the low-$n_s$ red sample exhibits a value of $A_\mathrm{IA}$ that is lower than this expectation by more than 1-$\sigma$. 

This finding suggests that the luminosity dependence observed for red galaxies is not universal, but it seems to be affected by the structure of galaxies. Red galaxies that are more disk-like and less concentrated in the centre seem to have a lower large-scale alignment amplitude that red galaxies that are more centrally concentrated. We note, however, that this difference is below 2$\sigma$. In addition, while we plot the logarithm of the average luminosity for each sample in Fig. \ref{fig:luminosityfit}, the red samples that are further sub-divided by S\'ersic index are not split by luminosity since we are limited by statistics. As such, these galaxy samples are not very localised in their luminosity, meaning very low luminosity galaxies could potentially be pulling the measured $A_\mathrm{IA}$ low for the red, low-$n_s$ sample. A larger sample of galaxies with higher statistical power is needed to be able to split them according to their colour, S\'ersic index, and luminosity, and further confirm this result. 

\subsection{IA for different radial weighting}
\label{sec:Results_weight}

\begin{figure}
    \centering
    \includegraphics[width=0.47\textwidth]{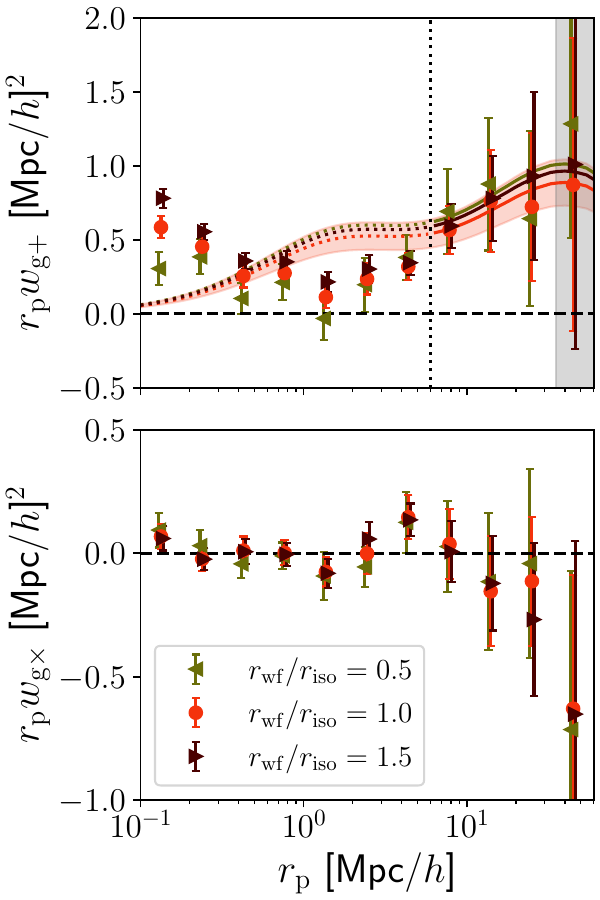}
    \caption{Similar to Fig. \ref{fig:luminosity},  but showing measurements of intrinsically red galaxies with shapes measured using different sizes for the radial weight function (compared the isophotal radius of galaxies), as indicated in the legend. Over-plotted in the top panel are the best-fit NLA models for each sub-sample, with a red band indicating the 1-$\sigma$ region around the best-fit model for the fiducial sample, with $r_\mathrm{wf}/r_\mathrm{iso}=1$.}
    \label{fig:rwf}
\end{figure}

When measuring galaxy shapes with a moment-based method, where a radial weight function needs to be applied to the galaxy images, it is possible to vary the size of the weight function to effectively probe the shape of different galaxy regions (see Sect. \ref{sec:Data_shapes}). A larger weight function will up-weight the outer regions of a galaxy compared to a smaller one. We used this to study galaxy alignments as a function of the scale at which the galaxy shape is measured. The size in our case is tied to the isophotal radius of the galaxy, and we measure it in ratios of $r_\mathrm{wf}/r_\mathrm{iso}$. 

To study the alignment signal as a function of galaxy scale, we first ensured we selected the exact same galaxies in all of the different shape-measurement catalogues by requiring that galaxies pass all flags for all shape measurements on top of any other galaxy sub-sample selection. This minimises any variation due to sub-sample selection between the different shape measurements. We then measured the $w_\mathrm{g+}$ correlation as before, and we present our results in Fig. \ref{fig:rwf}. 

As previously, we report that the $w_\mathrm{g\times}$ component is consistent with zero. The alignment measurements show a strong, non-zero signal for all methods across all scales. On large scales, the $w_\mathrm{g+}$ measured using the different shape methods do not differ significantly, but the values obtained using the largest weight function are higher than values using the smaller ones. Since the measurements are highly correlated, we quantified the significance of this difference by computing the quantity $w_\mathrm{g+}(\rp;r_\mathrm{wf}/r_\mathrm{iso}=1.5)-w_\mathrm{g+}(\rp;r_\mathrm{wf}/r_\mathrm{iso}=1.0)$ and its jackknife covariance. We chose this combination of values for the weight functions because it gives the highest S/N\footnote{Even though we measured shapes with a smaller weight function and we expect the signal to be stronger when using the largest weight function difference possible, the correlation we measure is noisier when using $r_\mathrm{wf}/r_\mathrm{iso}=0.5$. This can be due to the weight function being too small and susceptible to modelling bias. See \citet{Georgiou_DEIMOS} for more details.}. We find this is consistent with zero at 1$\sigma,$ and we fitted an NLA model to this $w_\mathrm{g+}$ difference between the two weight functions. Since the same galaxies were used here, the measurement is proportional to $b_g\,\Delta A_\mathrm{IA}$ with $\Delta A_\mathrm{IA}=A_\mathrm{IA}^{r_\mathrm{wf}/r_\mathrm{iso}=1.5}-A_\mathrm{IA}^{r_\mathrm{wf}/r_\mathrm{iso}=1.0}$, and we constrained the difference in large-scale alignment amplitude between the two shape measurement methods. We find $\Delta A_\mathrm{IA}=0.38\pm0.71$. Note that when fitting only to $w_\mathrm{g+}(\rp;r_\mathrm{wf}/r_\mathrm{iso}=1.0)$, we find $A_\mathrm{IA}^{r_\mathrm{wf}/r_\mathrm{iso}=1.0}=3.28\pm0.61$. We repeated this process for our highest S/N sample, which in Table \ref{tab:A_IA} is our highest luminosity intrinsically red galaxy sub-sample. We find similar results, with a $\Delta A_\mathrm{IA}=1.74\pm1.01$. This is a more significant result, which hints to the fact that this trend with increasing weight function size on large scales might be physical.

Interestingly, the picture is much more clear when looking at smaller scales, below 1 Mpc$/h$. The measurements show a clear trend, with the smallest and largest weight-function samples resulting in the lowest and highest measured alignment signals, respectively. While at small galaxy pair separations the effect of galaxy blending and stray light becomes important and can induce an artificial alignment signal, this effect is expected to be small, especially on scales of $\sim1$ Mpc$/h$ \citep[see e.g. Fig. 2 in][]{Georgiou_groups}. Hence, it is more likely that this difference in alignment is caused by isophotal twisting, making outer regions of satellite galaxies more strongly aligned than inner ones. We discuss the interpretation of these results further in Sect. \ref{sec:Conclusion}.

\subsection{IA of blue and disk-like galaxies}
\label{sec:Results_blue}

\begin{figure}
    \centering
    \includegraphics[width=0.47\textwidth]{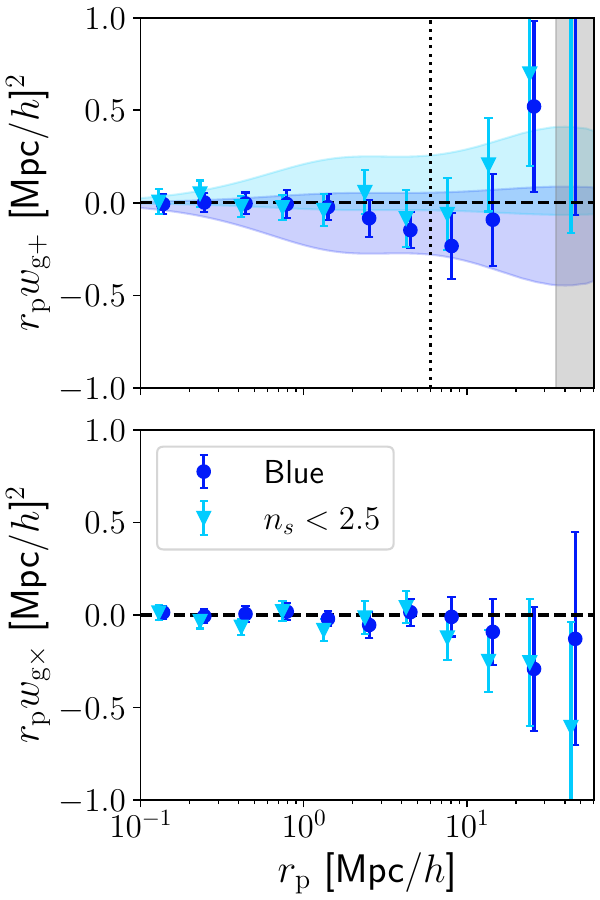}
    \caption{Similar to Fig. \ref{fig:luminosity}, but for the intrinsically blue and low-S\'ersic-index sub-samples with dark blue circles and light blue triangles, respectively. We also plot the 1-$\sigma$ region around the best-fit NLA models for both samples.}
    \label{fig:blue}
\end{figure}

Having looked at intrinsically red and/or elliptical galaxies, we turn our attention to blue and disk-like ($n_s<2.5$) galaxies in this section. For these two sub-samples, we performed the projected position--shape correlation function measurement similarly to in the previous sections and present the results in Fig. \ref{fig:blue}. Again, we note that the position--cross ellipticity measurements are consistent with zero. We see that the $w_\mathrm{g+}$ measurements are also very close to zero, meaning we do not detect a signal of either blue or low-$n_s$ galaxy alignment. We fitted the NLA model to these data points to determine bounds on the effective large-scale intrinsic alignment amplitude for these galaxies, and our best-fit values are presented in Table \ref{tab:A_IA}. These $A_\mathrm{IA}$ values are simply an effective description, since we do not expect the linear alignment model to accurately describe the alignment of rotationally supported galaxies, but its determination can be useful when mitigating the IA impact of these galaxies on weak lensing measurements.

\section{Conclusions}
\label{sec:Conclusion}

The aim of this work was to study the intrinsic alignment of galaxies and its relationship to several galaxy properties, such as intrinsic colour, luminosity, and structural parameters. 
We used the KiDS bright sample catalogue and matched with the structural parameter catalogue, which were derived using S\'ersic surface-brightness profile fits following \citet{Roy} and made available to the KiDS collaboration for internal use. We find that selecting intrinsically red galaxies or galaxies with a high S\'ersic index ($n_s>2.5$) does not significantly alter the power-law luminosity scaling of the large-scale intrinsic alignment amplitude. 

This luminosity dependence of intrinsic alignments is well-established in the literature \citep[e.g.][]{Joachimi, Singh15, KiDSLRG}. \citet{Joachimi} used a full sample to constrain $A_0=5.76^{+0.6}_{-0.62}$ and $\beta=1.13^{+0.25}_{-0.2}$, while the constraints of \citet{Singh15} were $A_0=4.9\pm0.6$ and $\beta=1.30\pm0.27$, both of which are in statistical agreement with our measurements. \citet{KiDSLRG} used a luminous red galaxy sample also based on KiDS data, and they measured $A_0=5.98\pm0.27$ and $\beta=0.93\pm0.11$, which is consistent with our results. That study also looked at using a double power law to describe the $A_\mathrm{IA}(L)$ function, which provided a better fit to the data and is likely a reflection of the 'knee' in the stellar-to-halo mass relation \citep{FortunaMass}. The S/N in our measurements is not high enough to be able to distinguish this feature, and for that reason we only constrained a single power-law luminosity dependence. 

When selecting red galaxies and then further sub-dividing them into high and low S\'ersic index (with a boundary at $n_s=4$), we see a hint that red, low-$n_s$ galaxies have a weaker alignment than expected from their average luminosity, though not at a high significance level. This galaxy sample is likely to have a large fraction of red spiral galaxies \citep{MRSG} that are rotationally supported and therefore expected to have a lower alignment signal, which would explain the trend seen in this work. If this behaviour is confirmed with better data in the future, it would change our basic understanding of how the alignment of intrinsically red galaxies depends on the galaxy's luminosity and mass. 

This may also have implications for the mitigation of intrinsic alignments in weak-lensing measurements. The galaxy S\'ersic index has been shown to evolve with redshift for very massive galaxies and is also a strong function of galaxy stellar mass \citep[e.g.][]{Sersic_z_Lang,Sersic_z_JWST}. In addition, the galaxy samples used across different redshift bins in weak-lensing analyses span a wide range of redshift and stellar mass. Consequently, using the same luminosity dependence on intrinsic alignment bias (e.g. keeping $A_0$ and $\beta$ the same) for galaxies in different redshift bins runs the risk of mismodelling intrinsic alignments, since galaxies in these bins can have a very different S\'ersic index distribution. It also means that if priors are determined for these parameters at low-redshift samples, they might not be accurate for the deep samples employed in weak-lensing measurements. Furthermore, our results can have implications for methods where cosmological inference is carried out using the connection between IA and the luminosity function \citep[see e.g.][]{Niko}. This motivates further exploration of the dependence of intrinsic alignment on the structural parameters of galaxies.

We also explored the dependence of the intrinsic alignment signal on the effective galaxy scale at which the shapes are measured. By varying the size of the weight function used in the shape measurement process, we are able to up-weight or down-weight outer regions of galaxies and measure their shape on different galaxy scales. At large galaxy pair separations, we do not see a statistically significant difference in the alignment signal, although our data are consistent with outer regions of a galaxy having an $\sim15-20\%$ stronger alignment signal (at 1-$\sigma$). At small separations, we see a clear galaxy-scale dependence, with outer regions of galaxies being much more strongly aligned than inner ones. 

Similar measurements have been made in the literature. In \citet{Singh16}, the NLA model is fitted to $w_\mathrm{g+}$ measurements from three different shape-measurement methods that each approaches galaxy-shape estimation in a different way. They find a difference of $\sim35\%$ in the largest and smallest estimated $A_\mathrm{IA}$, which is consistent with our findings. It is worth noting that these results are based on very different shape measurement methods that make it more complicated to compare them, since they depend on systematic effects in different ways, and it is less clear how the different regions of galaxies are weighted. In \citet{Georgiou_groups}, the alignment of satellite galaxies in GAMA galaxy groups was studied as a function of galaxy scale with the same shape-measurement method used in this work, and it was shown that outer regions of satellites align more strongly with their group's centre than inner ones. The alignment of satellite galaxies is expected to dominate the $w_\mathrm{g+}$ signal on small scales \citep{GAMAJohnston}, and we indeed see a stronger alignment on those scales for larger weight functions in Fig. \ref{fig:rwf}; this is in agreement with the measurements from \citet{Georgiou_groups}. 

Lastly, we also looked at intrinsically blue galaxies as well as galaxies with a low S\'ersic index ($n_s<2.5$). As these types of galaxies compose a large fraction of the samples used in weak-lensing measurements, it is interesting to explore their alignment signal. For both of these samples, we measure intrinsic alignments consistent with zero at the 1-$\sigma$ level, with bounds presented in Table \ref{tab:A_IA}. While constraints of emission-line galaxies exist in the literature, the flux-limited sample used here more closely resembles the galaxy samples used in weak-lensing measurements (though at much lower redshift) and can be used to better understand the level of contamination that can be expected in such weak-lensing measurements (extrapolating in redshift).

This is the first study that aims to measure the dependence of intrinsic alignment on galaxy structural parameters, which might reveal a more nuanced effect of tidal fields on galaxies than a simple intrinsic colour split. It motivates the creation of more surface-brightness profile measurements of galaxies that are typically used in direct intrinsic alignment studies. Such measurements can help gain a deeper understanding of the physical mechanisms with which galaxies align and improve our theoretical models and mitigation strategies.

\begin{acknowledgements}

    The authors thank Harry Johnston for making his projected correlation function measurement code \citep{GAMAJohnston} public\footnote{\url{https://github.com/harrysjohnston/2ptPipeline}}. 
    This publication is part of the project ``A rising tide: Galaxy intrinsic alignments as a new probe of cosmology and galaxy evolution'' (with project number VI.Vidi.203.011) of the Talent programme Vidi which is (partly) financed by the Dutch Research Council (NWO). This work is also part of the Delta ITP consortium, a program of the Netherlands Organisation for Scientific Research (NWO) that is funded by the Dutch Ministry of Education, Culture and Science (OCW). 
    CG is also supported by the MICINN project PID2022-141079NB-C32. 
    MB is supported by the Polish National Science Center through grants no. 2020/38/E/ST9/00395 and 2020/39/B/ST9/03494.
    NRN acknowledges support from the Guangdong Science Foundation grant (ID: 2022A1515012251).
    IFAE is partially funded by the CERCA program of the Generalitat de Catalunya.\\

    \emph{Author contributions:} All authors contributed to the development and writing of this paper. The authorship list is given in two groups: the lead authors (CG, NEC), followed by an alphabetical group which includes those who have either made a significant contribution to the data products or to the scientific analysis.\\

    Based on observations made with ESO Telescopes at the La Silla Paranal Observatory under programme IDs 177.A-3016, 177.A-3017, 177.A-3018 and 179.A-2004, and on data products produced by the KiDS consortium. The KiDS production team acknowledges support from: Deutsche Forschungsgemeinschaft, ERC, NOVA and NWO-M grants; Target; the University of Padova, and the University Federico II (Naples).\\

    GAMA is a joint European-Australasian project based around a spectroscopic campaign using the Anglo-Australian Telescope. The GAMA input catalogue is based on data taken from the Sloan Digital Sky Survey and the UKIRT Infrared Deep Sky Survey. Complementary imaging of the GAMA regions is being obtained by a number of independent survey programmes including GALEX MIS, VST KiDS, VISTA VIKING, WISE, Herschel-ATLAS, GMRT and ASKAP providing UV to radio coverage. GAMA is funded by the STFC (UK), the ARC (Australia), the AAO, and the participating institutions. The GAMA website is \url{https://www.gama-survey.org/}.
\end{acknowledgements}

\bibliographystyle{aa} 
\bibliography{references}

\begin{appendix}
\onecolumn

\section{Comparison of luminosity scaling with previous measurements}
\label{ap:literature}

\begin{figure*}[h!]
    \centering
    \includegraphics[width=0.70\textwidth]{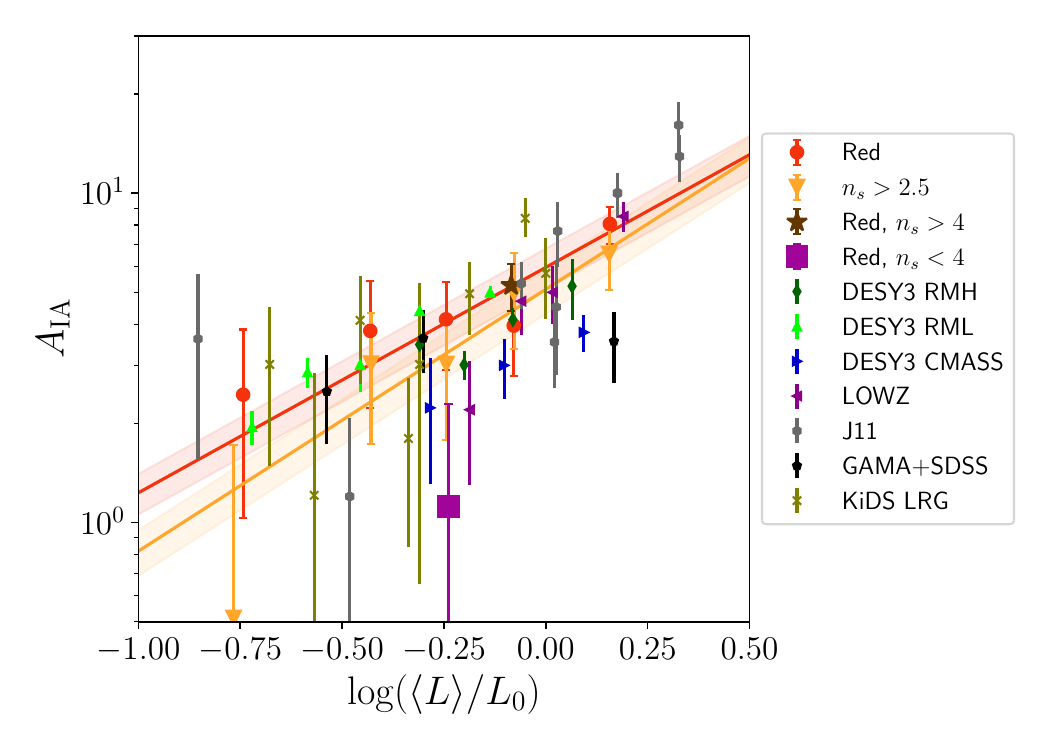}
    \caption{Same as Fig. \ref{fig:luminosityfit} but including measurements of $A_\mathrm{IA}$ from other studies: samples DESY3 RMH (dark green diamonds), DESY3 RML (light green upper triangles), and CMASS (blue right-facing triangles) are from \citet{Samuroff}, LOWZ (purple left-facing triangles) from \citet{Singh15}, J11 (gray hexagons) from \citet{Joachimi}, GAMA+SDSS (black pentagons) from \citet{GAMAJohnston} and KiDS LRG (olive x-markers) from \citet{Fortuna}. The last two samples are not entirely independent of our measurements in this work.}
    \label{fig:literature}
\end{figure*}

Here we compare our measurements of the large-scale intrinsic alignment amplitude, $A_\mathrm{IA}$, as a function of luminosity to previous measurements in the literature for red galaxy samples. The results are summarised in Fig. \ref{fig:literature}. Note that the galaxy samples used in each of the referenced measurements differ in selection, so exact agreement across datasets is not necessarily expected.

\section{Galaxy sample distributions}
\label{ap:distributions}

In this section we show distributions of several properties of our galaxy sample, which are summarised In Fig. \ref{ap:distributions}. The top panel of the figure shows the distribution of galaxy luminosity for red galaxies, as well as red galaxies split in S\'ersic index. The middle panel of the figure shows the distribution of S\'ersic index for the full galaxy sample, the intrinsically red, and blue galaxies. We note that blue galaxies have $n_s$ very localised around the value of 1 while red galaxies have a wider range of $n_s$ values. Lastly, the bottom panel of the figure shows the distribution of rest-frame $g-r$ colour for the full sample and for a split in S\'ersic index to distinguish between spheroidal ($n_s<2.5$) and elliptical ($n_s>2.5$) galaxies. Interestingly, we see that there is a population of spheroidal galaxies that exhibit red intrinsic colour. 

\begin{figure}[h!]
    \centering
    {\includegraphics[width=0.80\textwidth]{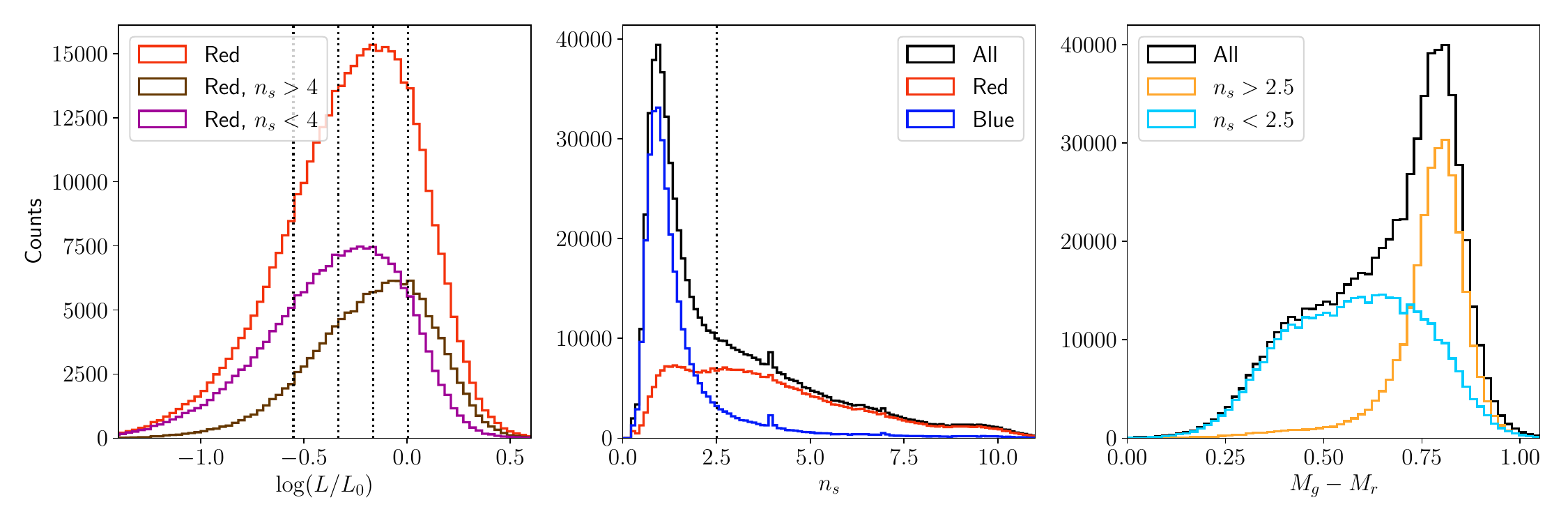}}
    \caption{Distributions of galaxy properties of our sample. \emph{Top}: Distributions of galaxy luminosity for our red sample (red line), red sample with high $n_s$ (brown), and red sample with low $n_s$ (purple). \emph{Middle}: Distributions of S\'ersic index, $n_s$, for the full sample, red, and blue galaxy samples (in black, red and blue line, respectively). \emph{Bottom}: Distributions of restframe $g-r$ colour for our full sample (black), spheroidal ($n_s<2.5$, light blue), and elliptical ($n_s>2.5$, orange) galaxy samples.}
    \label{fig:distributions}
\end{figure}

\end{appendix}

\end{document}